\documentclass[12pt]{article}
\usepackage{amsmath,amsfonts,amssymb,latexsym}

\numberwithin{equation}{section}
\def\be{\begin{equation}}
\def\ee{\end{equation}}
\def\bq{\begin{eqnarray}}
\def\eq{\end{eqnarray}}
\def\beq{\begin{eqnarray*}}
\def\eeq{\end{eqnarray*}}

\def\r{\rho}
\def\a{\alpha}
\def\b{\beta}
\def\g{\gamma}

\def\d{\delta}

\def\s{\sigma}

\def\z{\zeta}

\def\t{\theta}

\begin{document}
\title{\Huge{Asymptotic vacua with higher derivatives}}
\author{{\Large\textsc{Spiros Cotsakis\footnote{On leave from the University of the Aegean, 83200 Samos, Greece.}\,\,$^{1}$\thanks{\texttt{skot@aegean.gr}}, Seifedine Kadry$^{1}$\thanks{\texttt{Seifedine.Kadry@aum.edu.kw}},}}\\ {\Large\textsc{Giorgos Kolionis$^{2}$\thanks{\texttt{gkolionis@aegean.gr}}, Antonios Tsokaros$^{2}$\thanks{\texttt{atsok@aegean.gr}}}} \\
$^{1}$Department of Mathematics, \\ American University of the Middle East\\P. O. Box 220 Dasman, 15453, Kuwait \\
$^{2}$Research group of Geometry, Dynamical Systems and Cosmology,\\
University of the Aegean, Karlovassi 83200, Samos, Greece.}
\maketitle
\begin{abstract}
\noindent We study limits of vacuum, isotropic universes in the full, effective, four-dimensional theory with higher derivatives. We show that all flat vacua as well as general curved ones are globally attracted by the standard, square root scaling solution at early times. Open vacua asymptote to horizon-free, Milne states in both directions while closed universes exhibit more complex  logarithmic singularities, starting from initial data sets of a possibly smaller dimension. We also discuss the relation of our results to the asymptotic stability of the passage through the singularity in ekpyrotic and cyclic cosmologies.
\end{abstract}


\section{Introduction}
Vacuum states are not trivially obtainable for simple isotropic universes in general relativity, and one has to go beyond them to anisotropic, or more general inhomogeneous cosmologies for a vacuum to start making sense \cite{ycb}. However, isotropic vacua are very common in effective theories with higher derivatives, see e.g.,  \cite{star}-\cite{ba-mid2}. Such classical vacua are usually thought of as acquiring a physical significance when viewed as possible low-energy manifestations of a more fundamental superstring theory, although their treatment shows an intrinsic interest quite independently of the various quantum considerations.

In this paper, we consider the asymptotic limits towards  singularities of  vacuum universes coming from effective theories with higher derivatives. Such a study is  related to the existence and stability of an inflationary stage at early times in these contexts, and also to the intriguing possibilities of solutions with no particle horizons. For flat vacua, we find the general asymptotic solution with an early-time singularity. This result is then extended to cover general curved vacuum isotropic  solutions and we give the precise form of the attractor of all such universes with a past singularity. We also obtain special asymptotic states valid specifically for open or closed vacua starting from lower-dimensional initial data. These results have a potential importance for the ekpyrotic and cyclic scenarios as they strongly point to the dynamical stability of the reversal phase under higher derivative corrections in these universes.

The plan of this paper is as follows. In the next section, we derive the basic cosmological equations and show that they form an autonomous dynamical system in suitable variables. Then, in Sections 3-5, we apply asymptotic methods to study the early and late evolution of these isotropic cosmologies. In particular, we study separately the flat and curved subcases and show that there exist certain properties  valid asymptotically irrespective of the influence of curvature, while other  limits, coined here `Milne states',  have a very sensitive dependence on the sign of the constant curvature slices.


\section{The vacuum field}
Our starting point is a vacuum, FRW universe with a Robertson-Walker metric of the form
\be \label{rwmetrics}
g_{4}=-dt^{2}+a(t)^{2}\, g_{3},
\ee
where $a(t)$ is the scale factor.
Each slice is determined by the 3-metric
\be
g_{3}=\frac{1}{1-kr^2}dr^{2}+r^2\left(d\theta^{2}+\sin^{2}\theta d\phi^{2}\right),
\ee
where $k$ is the constant curvature normalized to take the values $0, +1$ or $-1$ for the complete, simply connected,  flat, closed or open space sections respectively. In what follows we use the sign conventions of \cite{mtw} and we set $8\pi G=c=1$.

Our general higher order action is
\be
\mathcal{S}=\frac{1}{2}\int_{\mathcal{M}^4}\mathcal{L}(R)d\mu_{g},\quad\mathcal{L}(R)=R + \b R^2 + \g \textrm{Ric}^2 + \d \textrm{Riem}^2 ,
\label{eq:lagra}
\ee
where $\b,\g,\d$ are constants. We consider a family of metrics $\{g_s:s\in\mathbb{R}\}$, and denote its compact variation by $\dot{g}_{\mu\nu}=(\partial g/\partial s)_{s=0}$. Using a $g$-variation of the action (\ref{eq:lagra}) to derive the field equations, only terms up to $\textrm{Ric}^2$ will occur, since in four dimensions we can use the Gauss-Bonnet identity,
\be
\dot{\mathcal{S}}_{GB}= \int_{\mathcal{M}^4}(R^2_{GB}d\mu_{g})^{\cdot}=0,\quad
R^2_{GB}=R^2 - 4\textrm{Ric}^2 + \textrm{Riem}^2,
\label{eq:gentity}
\ee
Below we focus in the case where $\mathcal{M}^4$ is a homogeneous and isotropic universe with metric (\ref{rwmetrics}), in which case we can also use the following identity,
\be
\int_{\mathcal{M}^4} ((R^2 - 3\textrm{Ric}^2)d\mu_{g})^{\cdot}=0 \:,
\label{eq:isontity}
\ee
Thus, by altering only the arbitrary constants we can further include the contribution of the $\textrm{Ric}^2$ term into the coefficient of $R^2$. The field equations will now read:
\be
R^{\mu\nu}-\frac{1}{2}g^{\mu\nu}R+
      \frac{\xi}{6} \left[2RR^{\mu\nu}-\frac{1}{2}R^2g^{\mu\nu}-2(g^{\mu\rho}g^{\nu\s}-g^{\mu\nu}g^{\rho\s})\nabla_{\rho}\nabla_{\s}R \right]=0,
\label{eq:fe}
\ee
where we have set  $\xi=2(2\b+\g+4\d-2)$. We note that because of the form of the coefficient $\xi$, some `memory' of the original fully quadratic theory (\ref{eq:lagra}) remains, and the final effective action leading to the field equations (\ref{eq:fe}) is not equivalent to an $R+\zeta R^2$ action with $\zeta$ \emph{arbitrary} (a use of the latter action would mean taking into account only the algebraic dependence of the quadratic curvature invariants).

Eq.(\ref{eq:fe}) splits into $00$- and $ii$-components ($i=1,2,3$) and it is well known \cite{bar-ott-83} that only the $00$-component is sufficient to form the field equation of our theory.
For the rest of this article an overdot will be used to denote differentiation with respect to the proper time, t. Using  the metric (\ref{rwmetrics}), the field equation (\ref{eq:fe}) takes the following form
\be
\frac{k+\dot{a}^2}{a^2}+\xi\left[2\: \frac{\dddot{a}\:\dot{a}}{a^2} + 2\:\frac{\ddot{a}\dot{a}^2}{a^3}-\frac{\ddot{a}^2}{a^2} - 3\:
\frac{\dot{a}^4}{a^4} -2k\frac{\dot{a}^2}{a^4} + \frac{k^2}{a^4}\right] = 0.
\label{eq:fe2}
\ee
This equation has identical left hand side to the corresponding equation of the radiation problem studied in \cite{kolionis1} (see Eq. (2.10) in that reference), while its right hand side can be obtained from that of the latter, namely, ${\zeta^2}/{a^4}$  ($\zeta$ is a constant satisfying the constraint $\rho=3\zeta^2/a^4$ valid for the radiation problem of Ref. \cite{kolionis1}), by letting the constant $\zeta$ tend to zero. So the first question we like to address is: Why not let $\zeta$ tend to zero also in the \emph{solutions} of the radiation problem found in \cite{kolionis1} to directly obtain the solution of the vacuum problem considered in the present work? Actually, as we shall see below, the answer to this question is related to a subtle property of possible vaccum solutions, one that is not obtainable when considering the naive asymptotic limit of letting the matter terms tend to zero in the dynamical equations.

In \cite{kolionis1}, we showed that all radiation asymptotics come from the unique acceptable asymptotic decomposition of the radiation problem considered in that work. Namely, the only possible  asymptotic splitting of the vector field  $\mathbf{f}^{(0)}_{\,k,\textsc{RAD}}$ on approach to the spacetime singularity is given by,
\be
\mathbf{f}_{\,k,\textsc{RAD}}=\mathbf{f}^{(0)}_{\,k,\textsc{RAD}} + \mathbf{f}^{\,(\textrm{sub})}_{\,k,\textsc{RAD}},
\ee
with dominant part
\be
\mathbf{f}^{(0)}_{\,k,\textsc{RAD}}(\mathbf{x})=\left(y,z,\frac{3y^3}{2x^2} + \frac{z^2}{2y} -\frac{yz}{x}
\right), \label{eq:f0}
\ee
and subdominant part
\be\label{eq:f01}
\mathbf{f}^{\,(\textrm{sub})}_{\,k,\textsc{RAD}}(\mathbf{x})=
\left(0,0, \frac{\z^2-k^2\xi}{2\xi x^2y} - \frac{y}{2\xi} -\frac{k}{2\xi y} + \frac{ky}{x^2}\right).
\ee
From this asymptotic decomposition it follows that all terms involving possible vacuum limits, i.e., terms that include the parameter $\zeta$, as well as the curvature terms having the constant $k$, appear only in the \emph{subdominant} part asymptotically, cf. Eq. (\ref{eq:f01}). In particular, they are totally absent from the dominant part of the vector field splitting that is from Eq. (\ref{eq:f0}).

Therefore the set of vacuum solutions obtained in this way, that is by letting the constant $\zeta$ tend to zero in any radiation solution,  cannot obviously exhaust all dynamical possibilities of vacuum evolution as it misses precisely those vacuum solutions in which the relevant terms enter dominantly. It also shows that the choice of the unknowns in the analysis of the radiation problem (namely,  $x=a$, $y=\dot{a}$ and $z=\ddot{a}$) is suitable only for that problem.

In the present work, we are interested in tracing all possible vacuum asymptotics, especially those solutions for which curvature and  vacuum enter in the \emph{dominant} part of the vector field asymptotically. We expect that in terms of suitable new variables, the vacuum problem will show certain asymptotes not obtainable from the radiation problem when letting   the constant $\zeta$ tend to zero, and these will only be possible in decompositions allowing the curvature and the terms characterizing the vacuum be present in the dominant part of the field asymptotically (something impossible in the radiation problem). We also expect that other decompositions in the new variables will lead to solutions which converge to those obtained from the radiation ones  by letting the constant $\zeta$ tend to zero, and these will exactly correspond to,  and stem from,  decompositions having the curvature and vacuum terms  only in the subdominant part asymptotically.

As it turns out both of these expectations are indeed realized. Setting $H=\dot{a}/a$ for the Hubble expansion rate, Eq. (\ref{eq:fe2}) reads,
\be \ddot{H}=\frac{1}{2}\frac{\dot{H}^2}{H}-3H\dot{H}+\frac{k}{a^2}H-\frac{1}{2}\frac{k^2}{a^4}\frac{1}{H}
-\frac{1}{12\epsilon}H-\frac{k}{12\epsilon a^2}\frac{1}{H}
\label{eq:beq}
\ee
where now we have put $\epsilon =\xi / 6$. We then introduce new variables for the present problem by setting
 $x=H$, $y=\dot{H}$ and $z=a^{-2}$, so that Eq. (\ref{eq:beq}) can be written as an autonomous dynamical
system of the general form
\be\label{basic dynamical system}
\mathbf{\dot{x}}=\mathbf{f}_{\textsc{VAC}}(\mathbf{x}),\quad \mathbf{x}=(x,y,z),
\ee
that is
\begin{eqnarray}
\label{eq:ds}
\dot{x} &=& y \:\:\:\:\:\nonumber \\
\dot{y} &=& \frac{y^{2}}{2x}-3xy+kxz-\frac{k^2 z^2}{2x}- \frac{x}{12\epsilon}-\frac{kz}{12\epsilon x}\:\:\:\:\:\label{basic dynamical system1}  \\
\dot{z} &=& -2xz\:\:\:\:\:\nonumber
\end{eqnarray}
 equivalent to the \emph{vacuum}, 3-dimensional  vector field $\mathbf{f}_{\textsc{VAC}}:\mathbb{R}^3\rightarrow\mathbb{R}^3$ with
\be\label{vf}
\mathbf{f}_{\textsc{VAC}}(x,y,z)=\left( y,\frac{y^{2}}{2x}-3xy+kxz-\frac{k^2 z^2}{2x}- \frac{x}{12\epsilon}-\frac{kz}{12\epsilon x},-2xz \right).
\ee
This  field  completely describes the dynamical evolution of a vacuum, flat or curved, FRW universe in the gravity theory defined by the full quadratic action (\ref{eq:lagra}). We shall assume that $x\neq 0$, that is we consider only non-static universes. In the following two sections, we consider the case when the curvature and vacuum terms enter the decomposition subdominantly. As shown at the end of Section 3 and 4, the found solutions (in the new variables) indeed correspond to the radiation ones  by letting the constant $\zeta$ tend to zero, meaning that these forms are indeed possible in the general vacuum evolution. In the curved case this comes from the asymptotic splitting $\mathbf{f}^{(0)}_{\,k,\textsc{VAC}_1}$. However, the vacuum field has more decompositions, namely $\mathbf{f}^{(0)}_{\,k,\textsc{VAC}_2}$ and $\mathbf{f}^{(0)}_{\,k,\textsc{VAC}_3}$, that now include the effects of vacuum and curvature appearing in their dominant part asymptotically, cf. beginning of Section 4 and Section 5, impossible in the radiation problem. Using these forms, we are able to find new asymptotic vacua not having any relation to those obtained from the radiation ones  by letting the constant $\zeta$ tend to zero. These in turn lead to Milne type attractors monitoring precisely the dominant effects of vacuum and curvature in the asymptotic evolution.

Let us end this section with the following remark about the general problem of having a fluid with a general equation of state $p=w\rho$, instead of pure radiation or vacuum. In this case, new terms would appear, for instance  the form
$$
\left(y,z,\frac{\z^2}{2\xi x^{3w+1}y}\right),
$$
in place of simple radiation terms (cf. Eq. (3.5) in \cite{kolionis1}). Although in the limits of radiation and vacuum, this reduces to the known forms, it is a new problem altogether since one needs to consider all different ranges of values of the fluid parameter $w$ to see if new forms of asymptotic evolution are possible.


\section{The unique flat vacuum }
When $k=0$, the vacuum field (\ref{vf}) becomes
\be\label{fvf}
\mathbf{f}_{\,0,\textsc{VAC}}(x,y,z)=\left( y, \frac{y^2}{2x}-3xy-\frac{x}{12\epsilon}, -2xz \right),
\ee
and the system (\ref{basic dynamical system1}) reads,
\begin{eqnarray}
\label{eq:fds}
\dot{x} &=& y \nonumber \\
\dot{y} &=& \frac{y^{2}}{2x}-3xy- \frac{x}{12\epsilon}\label{flat dynamical system}  \\
\dot{z} &=& -2xz\nonumber
\end{eqnarray}
Our main interest below is to study the behavior of the universe described by (\ref{fvf}), (\ref{flat dynamical system}) near the initial singularity, taken here to lie at $t=0$ (it is really arbitrary, however, and we could have placed it at any $t_0$ and used the variable $\tau=t-t_0$ instead of $t$). Following the method of asymptotic splittings of Refs.  \cite{CB,go,me12}, we find that of the $2^3-1=7$ possible asymptotic  decompositions that the field (\ref{fvf}) possesses, there is only one that  leads to a fully acceptable dominant balance, namely,
\be\label{basic dec}
\mathbf{f}_{\,0,\textsc{VAC}}=\mathbf{f}^{(0)}_{\,0,\textsc{VAC}} + \mathbf{f}^{\,(\textrm{sub})}_{\,0,\textsc{VAC}},
\ee
with dominant part
\be
\mathbf{f}^{(0)}_{\,0,\textsc{VAC}}(\mathbf{x})=\left(y, \frac{y^2}{2x}-3xy, -2xz \right), \label{eq:flatdom}
\ee
and subdominant part
\be\label{eq:flatsub}
\mathbf{f}^{\,(\textrm{sub})}_{\,0,\textsc{VAC}}(\mathbf{x})=
\left(0, - \frac{x}{12\epsilon}, 0 \right).
\ee
We shall  construct series solutions which encode information about the leading order behavior of all possible modes of asymptotic evolution,  as well as their generality (number of arbitrary constants) near the spacetime singularity at $t=0$. We recall that for any given dominant asymptotic solution of the system (\ref{eq:fds}),  the pair $(\mathbf{a},\mathbf{p})$ is a \emph{dominant balance} of the vector field $\mathbf{f}_{\,0,\textsc{VAC}}$, where
$\mathbf{a}=(\theta, \eta, \rho)\in\mathbb{C}^3$ are constants and $\mathbf{p}=(p, q, r)\in\mathbb{Q}^3$, and we look for a leading order behavior of the form
\be
\mathbf{x}(t)=\mathbf{a}t^{\mathbf{p}}=(\theta t^{p}, \eta
t^{q}, \rho t^{r}). \label{eq:domisol}
\ee
Such behaviors correspond to the asymptotic forms of the integral curves  of the vacuum field $\mathbf{f}_{\,0,\textsc{VAC}}$, as we take it to a neighborhood of the singularity.

Substituting the forms (\ref{eq:domisol}) into the dominant system $(\dot x,\dot y,\dot z)(t)=\mathbf{f}^{(0)}_{\,0,\textsc{VAC}}$ and solving the resulting nonlinear algebraic system to determine  the dominant balance $(\mathbf{a},\mathbf{p})$ as an exact, scale invariant solution, leads to the unique \emph{flat-vacuum} balance, denoted here by $\mathcal{B}_{\,0,\textsc{VAC}}\in\mathbb{C}^3\times\mathbb{Q}^3$,  with
\be\label{flat balance}
\mathcal{B}_{\,0,\textsc{VAC}}=(\mathbf{a},\mathbf{p})= \left(\left(
\frac{1}{2},-\frac{1}{2},\rho \right),\:
\left(-1,-2,-1\right)\right).
\ee
In particular, this means that the vector field $\mathbf{f}^{(0)}_{\,0,\textsc{VAC}}$ is a \emph{scale-invariant system}, cf. \cite{CB,go,me12}.

Further, we need to show that the term (\ref{eq:flatsub}) in the basic decomposition (\ref{basic dec}) of the flat-vacuum field (\ref{fvf}) is itself weight-homogeneous with respect to the flat-vacuum balance (\ref{flat balance}), for this splitting to be  acceptable. For this, we need to check that this \emph{candidate} subdominant part is indeed subdominant. Using the balance $\mathcal{B}_{\,0,\textsc{VAC}} $ of Eq. (\ref{flat balance}), we find that
\be\label{eq:fsub0vac}
\frac{\mathbf{f}^{(\textrm{sub})}_{\,0,\textsc{VAC}}(\mathbf{a}t^{\mathbf{p}})}{t^{\mathbf{p}-1}}
=\mathbf{f}^{(\textrm{sub})}_{\,0,\textsc{VAC}}(\mathbf{a})\,t^2
=\left(0, - \frac{t^2}{24\epsilon}, 0 \right),
\ee
and since $\mathbf{f}^{(sub)}_{\,0,\textsc{VAC}}(\mathbf{a})$ is different from zero, we find that when $\epsilon \neq 0$, that is for all cases except when $2\b+\g+4\d-2 =0$, the right-hand-side in Eq. (\ref{eq:fsub0vac}) goes to zero asymptotically when  $t\rightarrow 0$. Thus, we can safely conclude that the decomposition (\ref{basic dec}) is acceptable asymptotically  in every higher-order gravity theory when this constraint holds true\footnote{A physical example that corresponds to the choice $\epsilon=0$ and  is therefore excluded in this analysis is the conformally invariant Bach-Weyl gravity cf. \cite{bach+weyl}. Note that the same constraint appears in the stability analysis of the purely radiation universes in these theories, cf. Ref. \cite{kolionis1}.}.

We then find that in the present case the Kovalevskaya matrix corresponding to this solution is given by,
\be
\mathcal{K}_{\,0,\textsc{VAC}}=\left(
                     \begin{array}{ccc}
                       1 & 1   & 0\\
                       1 & -1/2 & 0 \\
                       -2\r&0   & 0
                     \end{array}
                   \right),
\ee
with spectrum
\be
\textrm{spec}(\mathcal{K}_{\,0,\textsc{VAC}})=\{-1,0,3/2\}.
\ee
Since the $\textrm{spec}(\mathcal{K}_{\,0,\textsc{VAC}})$ in our case possesses two non-negative eigenvalues, the balance $\mathcal{B}_{\,0,\textsc{VAC}}$ indeed corresponds to the dominant behaviour of a \emph{general} solution having the form of a formal series and valid locally around the initial singularity. To find it, we substitute the \emph{Fuchsian series expansions} (no constant first term, rational exponents)
\begin{equation} \label{eq:flatseries}
x(t) = \sum_{i=0}^{\infty} c_{1i} t^{\frac{i}{2}-1}, \:\:\:\:\:
y(t) = \sum_{i=0}^{\infty} c_{2i}  t^{\frac{i}{2}-2},\:\:\:\:\:
z(t) = \sum_{i=0}^{\infty} c_{3i} t^{\frac{i}{2}-1} ,\:\:\:\:\:
\end{equation}
where  $c_{10}=1/2 ,\:\:c_{20}=-1/2 ,\:\:c_{30}=\r\:\:$, in the original system (\ref{eq:fds}) and we are led to various recursion relations that determine the unknowns $c_{1i}, c_{2i}, c_{3i}$ term by term. The final result is,
\be
x(t) = \frac{1}{2}\:\:t^{-1} +c_{13}\:\:t^{1/2} -
\frac{1}{36\epsilon}\:\: t + \cdots .
\label{eq:flatgensol}
\ee
The corresponding series expansion for $y(t)$ is given by  the first time derivative of the above expression, while the corresponding series expansion for $z(t)$ is given by
\be
z(t) = \r\:\:t^{-1} -\frac{4\r \: c_{13}}{3} \:\:t^{1/2} +\frac{1}{36\epsilon} \:\: t + \cdots .
\label{eq:flatsolz}
\ee
Finally, we arrive at the following asymptotic form of the scale factor around
the singularity:
\be \label{eq:flatfinal}
a(t) =\a\:\:t^{1/2}+\frac{2c_{13}\a}{3}\:\:t^{2}-
\frac{\a}{72\epsilon}\:\:t^{5/2}+\frac{4\a\: c_{13}^{2}}{9}\:\:t^{7/2}+\cdots, \ee
where $\a$ is a constant of integration and $\a^{-2}=\r$.

As a final test we use the Fredholm's alternative for admission of this solution. This leads to a \emph{compatibility condition} for the positive eigenvalue 3/2 and the associated eigenvector at the $j=2$ level,
\be (2,1,-\frac{8\r}{3}) \cdot \left( \begin{array}{l}
                              -\frac{1}{2}c_{13}+c_{23}  \\
                                 c_{13}-2c_{23}   \\
                               -2\r c_{13}-\frac{3}{2}c_{33}
                                 \end{array}
              \right) = 0,
\label{eq:cc}
\ee
which is indeed  satisfied, thus leading to the conclusion that (\ref{eq:flatgensol})-(\ref{eq:flatsolz}) corresponds  to a valid asymptotic solution around the singularity.

Our series solution (\ref{eq:flatgensol})-(\ref{eq:flatsolz}) has three arbitrary constants, $c_{13}$, $\rho$ and another one corresponding to the arbitrary position of the singularity (taken here to be zero without loss of generality), and so  there is an open set of initial conditions  for which the general solution blows up at the finite time (initial) singularity at $t=0$.  This proves the stability of our solution in the neighborhood of the singularity\footnote{We can also see that the transformation $c_{13}={3c^{'}_{13}/2\a}$ and $\epsilon = -k/6 $ in the series expansion (\ref{eq:flatfinal}) leads to a form  obtained by setting $\z=0$ in the series expansion found for the flat, radiation case of Ref. \cite{K1} (cf. Eq. (21) in that reference, where the term $12\xi\t^3$ was mistakenly written as $24\xi\t^3$ there).}.


\section{Curved vacua}
As saw in the previous Section, in a flat, vacuum FRW model in the fully quadratic theory of gravity defined by the action (\ref{eq:lagra}), the vector field $\mathbf{f}_{\,0,\textsc{VAC}}$ has only one admissible asymptotic solution near the initial singularity, namely, the form (\ref{eq:flatgensol})-(\ref{eq:flatsolz}). In this family, all flat vacua are asymptotically dominated (or `attracted') at early times by the form $a(t)\sim t^{1/2}$, thus proving the stability of this solution in the flat case.

When $k\neq 0$, and we consider the situation of a vacuum but curved family of FRW universes, the vacuum field $\mathbf{f}_{\textsc{VAC}}$ has more terms than those present in the flat case, namely, those that contain $k$ in (\ref{vf}). Below we shall use the suggestive notation $\mathbf{f}_{\,k,\textsc{VAC}}$ instead of $\mathbf{f}_{\textsc{VAC}}$  to  distinguish spherical from hyperbolic (open) vacua. The field $\mathbf{f}_{\,k,\textsc{VAC}}$ (or the basic system (\ref{eq:ds})) can decompose precisely in $2^6-1=63$
different ways. Of these 63 decompositions, there are only three that eventually lead to fully acceptable dominant balances, while the rest 60 decompositions fail to lead to an acceptable picture for various different asymptotic reasons.
The acceptable asymptotic splittings of the vector field
$
\mathbf{f}_{\,k,\textsc{VAC}}=\mathbf{f}^{(0)}_{\,k,\textsc{VAC}} + \mathbf{f}^{\,(\textrm{sub})}_{\,k,\textsc{VAC}},
$
have dominant parts
\be \label{eq:curveddom1}
\mathbf{f}^{(0)}_{\,k,\textsc{VAC}_1} =\left(y, \frac{y^2}{2x}-3xy,-2xz\right) ,
\ee
\be \label{eq:curveddom2}
\mathbf{f}^{(0)}_{\,k,\textsc{VAC}_{2}}=\left(y, -3xy+kxz,-2xz\right) ,
\ee
\be \label{eq:curveddom3}
\mathbf{f}^{(0)}_{\,k,\textsc{VAC}_{3}}=\left(y, \frac{y^2}{2x}-3xy+kxz-\frac{k^2 z^2}{2x},-2xz\right) ,
\ee
while their subdominant parts are given respectively by the forms,
\be \label{eq:curvedsub1}
\mathbf{f}^{(\textrm{sub})}_{\,k,\textsc{VAC}_1} =\left(0, kxz-\frac{k^2 z^2}{2x}-\frac{x}{12\epsilon}-\frac{kz}{12x\epsilon},0\right) ,
\ee
\be \label{eq:curvedsub2}
\mathbf{f}^{(\textrm{sub})}_{\,k,\textsc{VAC}_2}=\left(0, \frac{y^2}{2x}-\frac{k^2 z^2}{2x}-\frac{x}{12\epsilon}-\frac{kz}{12x\epsilon},0\right) ,
\ee
\be \label{eq:curvedsub3}
\mathbf{f}^{(\textrm{sub})}_{\,k,\textsc{VAC}_3}=\left(0, -\frac{x}{12\epsilon}-\frac{kz}{12x\epsilon},0\right).
\ee
The first decomposition (\ref{eq:curveddom1}) has identical dominant part as the flat splitting of the previous section, hence identical dominant balance, namely, its asymptotic balance is $\mathcal{B}_{\,k,\textsc{VAC}_1} \in\mathbb{C}^3\times\mathbb{Q}^3$,  with
\be \label{curvedbalance1}
\mathcal{B}_{\,k,\textsc{VAC}_1}=(\mathbf{a},\mathbf{p})= \left(\left(\frac{1}{2},-\frac{1}{2},\r \right),\: \left(-1,-2,-1\right)\right),
\ee
In particular, this means that the vector field $\mathbf{f}^{(0)}_{\,k,\textsc{VAC}_1}$ is a scale-invariant system. However, its subdominant part (\ref{eq:curvedsub1}) is different, and  we need to show that   the higher-order terms (\ref{eq:curvedsub1}) in the basic decomposition of the vacuum field  are themselves weight-homogeneous with respect to the balance (\ref{curvedbalance1}), for admissibility. To prove this, we first split the subdominant part (\ref{eq:curvedsub1})  by writing
\be
\mathbf{f}^{\,(\textrm{sub})}_{\,k,\textsc{VAC}_1}(\mathbf{x}) = \mathbf{f}^{(1)}_{\,k,\textsc{VAC}_1}(\mathbf{x}) +
\mathbf{f}^{(2)}_{\,k,\textsc{VAC}_1}(\mathbf{x}) + \mathbf{f}^{(3)}_{\,k,\textsc{VAC}_1}(\mathbf{x}),
\ee
where
\be
\mathbf{f}^{(1)}_{\,k,\textsc{VAC}_1}(\mathbf{x})=\left(0,kxz,0 \right),\,\,
\mathbf{f}^{(2)}_{\,k,\textsc{VAC}_1}(\mathbf{x})=\left(0,- \frac{k^2 z^2}{2x}-\frac{x}{12\epsilon},0 \right),\,\,
\mathbf{f}^{(3)}_{\,k,\textsc{VAC}_1}(\mathbf{x})=\left(0,-\frac{kz}{12x\epsilon},0 \right),
\ee
and using the balance $\mathcal{B}_{\,k,\textsc{VAC}_1} $ defined by Eq. (\ref{curvedbalance1}), we find that
\bq
\frac{\mathbf{f}^{(1)}_{\,k,\textsc{VAC}_1}(\mathbf{a}t^{\mathbf{p}})}{t^{\mathbf{p}-1}}&=&
\mathbf{f}^{(1)}_{\,k,\textsc{VAC}_1}(\mathbf{a})t =\left(0,\frac{k\:\r}{2}\:\:t,0 \right),\\ \frac{\mathbf{f}^{(2)}_{\,k,\textsc{VAC}_1}(\mathbf{a}t^{\mathbf{p}})}{t^{\mathbf{p}-1}}&=&
\mathbf{f}^{(2)}_{\,k,\textsc{VAC}_1}(\mathbf{a})t^{2}=\left(0,\left(- k^2 \r^2-\frac{1}{24\epsilon}\right)\:\:t^2,0 \right),\\ \frac{\mathbf{f}^{(3)}_{\,k,\textsc{VAC}_1}(\mathbf{a}t^{\mathbf{p}})}{t^{\mathbf{p}-1}}&=&
\mathbf{f}^{(3)}_{\,k,\textsc{VAC}_1}(\mathbf{a})t^{3}=\left(0,-\frac{k\r}{6\epsilon}\:\:t^3,0 \right).
\eq
Hence, taking the limit as $t\rightarrow 0$, we see that these forms go to zero asymptotically provided that $\mathbf{f}^{(i)}_{\,k,\textsc{VAC}_1}(\mathbf{a}), i=1,2,3$ are  all different from zero, which happens only when
$\epsilon\neq 0$.
Since the \emph{subdominant exponents}
\be\label{sub exps}
q^{(0)}=0\ <\ q^{(1)}=1\ <\ q^{(2)}=2\ <\ q^{(3)}=3,
\ee
are ordered, we conclude that the subdominant part (\ref{eq:curvedsub1}) is weight-homogeneous as promised. Further, since the Kovalevskaya matrix and its spectrum are identical to the flat vacuum case, we arrive at the following asymptotic series representation for the decomposition (\ref{eq:curveddom1}):
\be
x(t) = \frac{1}{2}\:\:t^{-1} -\frac{k\r}{2} + c_{13}\:\:t^{1/2} - \left(\frac{k^2 \r^2}{4}+\frac{1}{36\epsilon} \right)\:\: t + \cdots,
\label{eq:gensol}
\ee
while the corresponding series expansion for $y(t)$ is given by  the first time derivative of the above expression, and that for $z(t)$ is given by
\be
z(t) = \r\:\:t^{-1} -k\r^2 -\frac{4\r \: c_{13}}{3} \:\:t^{1/2} + \left(\frac{k^2 \r^2 (1+2\r)}{4}+\frac{1}{36\epsilon} \right)\:\: t + \cdots .
\label{eq:gensolz}
\ee
For the scale factor, we find
\be \label{eq:curvedfinalgen}
a(t) =\a\:\:t^{1/2}-\frac{k\r \a}{2}\:\:t^{3/2}+\frac{2c_{13}\a}{3}\:\:t^{2}-\left( \frac{k^2 \r^2 \a}{8}+\frac{\a}{72\epsilon} \right)\:\:t^{5/2}+\cdots,
\ee
where $\a$ is a constant of integration and $\a^{-2}=\r$. This series (\ref{eq:gensol}) has two arbitrary constants, $\r, c_{13}$ and a third one corresponding to the arbitrary position of the singularity, and is therefore a local expansion of the \emph{general} solution around the initial singularity. The transformation $c_{13}={3c^{'}_{13}/2\a}$ and $\epsilon = k/6 $ in the series expansion (\ref{eq:curvedfinalgen}) leads to the form which is obtained by setting $\z=0$ in the series expansion found for the curved, radiation case, cf. Eq. (4.13) of \cite{kolionis1}. In addition, by setting $k=0$ we are lead to the form (\ref{eq:flatgensol}) found for the flat vacuum.

We note that because of the square root, limits can only be taken in the backward direction, $t\rightarrow 0$, in the solution (\ref{eq:curvedfinalgen}), another way of expressing the curious fact that this solution (along with Eq. (\ref{eq:flatfinal}) found in the previous Section) is \emph{only} valid at early times and corresponds to a past singularity.


\section{Milne states}
We now move on to the analysis of the last two decompositions, namely, those with dominant parts (\ref{eq:curveddom2}) and (\ref{eq:curveddom3}). We show below that these lead to particular solutions for $k=-1$ and $k=+1$. In the case of open universes,  $k=-1$, and the dominant parts take the forms
\be \label{eq:curveddom2 -1}
\mathbf{f}^{(0)}_{\,-1,\textsc{VAC}_{2}}=\left(y, -3xy-xz, -2xz\right) , \ee \be \label{eq:curveddom3 -1} \mathbf{f}^{(0)}_{\,-1,\textsc{VAC}_{3}}=\left(y, \frac{y^2}{2x}-3xy-xz-\frac{z^2}{2x}, -2xz\right) ,
\ee
with subdominant parts given by
\be \label{eq:curvedsub2 -1} \mathbf{f}^{(sub)}_{\,-1,\textsc{VAC}_2}=\left(0, \frac{y^2}{2x}-\frac{z^2}{2x}-\frac{x}{12\epsilon}+\frac{z}{12x\epsilon},0\right) ,
\ee
\be \label{eq:curvedsub3 -1} \mathbf{f}^{(sub)}_{\,-1,\textsc{VAC}_3}=\left(0, -\frac{x}{12\epsilon}+\frac{z}{12x\epsilon},0\right) ,
\ee
respectively. These two forms  lead to the same acceptable asymptotic balance
\be \label{curvedbalance2 -1} \mathcal{B}_{\,-1,\textsc{VAC}_{2,3}}=(\mathbf{a},\mathbf{p})= \left(\left(1,-1,1 \right),\: \left(-1,-2,-2\right)\right),
\ee
while the structure of the  $\mathcal{K}$-matrices is
\be
\mathcal{K}_{\,-1,\textsc{VAC}_2}=\left(
                     \begin{array}{ccc}
                       1 & 1   & 0\\
                       2 & -1 & -1 \\
                       -2 & 0  & 0
                     \end{array}
                   \right),\quad\textrm{spec}(\mathcal{K}_{\,-1,\textsc{VAC}_2})=\{-1,-1,2\},
\ee
and
\be
\mathcal{K}_{\,-1,\textsc{VAC}_3}=\left(
                     \begin{array}{ccc}
                       1 & 1   & 0\\
                       2 & -2 & -2 \\
                       -2 &0   & 0
                     \end{array}
                   \right),\quad\textrm{spec}(\mathcal{K}_{\,-1,\textsc{VAC}_3})=\{-1,-2,2\}.
\ee
Since we are interested in the behavior of solutions near singularities, we set the arbitrary constants corresponding to the negative eigenvalues equal to zero, and we are led to seeking only for particular solutions (one constant less that general ones). We find the following form for $x(t)$, common for both decompositions,
\be \label{eq:parsol1}
x(t) =t^{-1}+c_{12}\:\:t-\left(\frac{c_{12}-18\: \epsilon \: c_{12}^2}{60\epsilon}\right) \:\: t^3+\cdots . \ee
The corresponding series expansion for $y(t)$ is given by the first time derivative of the above expression, while the corresponding series expansion for $z(t)$ is given by \be
z(t) = t^{-2} -c_{12} + \left(\frac{c_{12} \:(\:42\epsilon \:c_{12}+1\:)}{120 \epsilon}\right)\:\: t^2 + \cdots . \label{eq:parsol1z} \ee
Finally, for the scale factor we find  the asymptotic expansion,
\be
a(t)=\alpha\:t+\frac{\alpha\: c_{12}}{2}\:\:t^3-\frac{\alpha \left(c_{12}-18\:\epsilon \: c_{12}^2 \right)}{240\epsilon}\:\:t^5 + \cdots , \ee where  $\alpha =\pm1$ as dictated by the definition $z(t)=1/a(t)^2$.
This solution  represents  a 2-parameter family of  past, or  future Milne states for these open vacua. It is  reminiscent of the Frenkel-Brecher horizonless solutions \cite{horizonless1}, with the important difference that their solutions are matter-filled an possibly valid only in the past direction.

On the other hand, when $k=+1$,  the decomposition  (\ref{eq:curveddom2}) does not lead to an acceptable dominant balance, but (\ref{eq:curveddom3}) does, namely,
\be \label{eq:curveddom3 +1} \mathbf{f}^{(0)}_{\,+1,\textsc{VAC}_{3}}=\left(y, \frac{y^2}{2x}-3xy+xz-\frac{z^2}{2x}, -2xz\right) , \ee with subdominant part \be \label{eq:curvedsub3 +1} \mathbf{f}^{(sub)}_{\,+1,\textsc{VAC}_3}=\left(0, -\frac{x}{12\epsilon}-\frac{z}{12x\epsilon}, 0\right) ,
\ee
 and we obtain \be \label{curvedbalance2 +1} \mathcal{B}_{\,+1,\textsc{VAC}_3}=(\mathbf{a},\mathbf{p})= \left(\left(1,-1,3 \right),\: \left(-1,-2,-2\right)\right).
\ee
The corresponding $\mathcal{K}$-matrix is

\be
\mathcal{K}_{\,+1,\textsc{VAC}_3}=\left(
                     \begin{array}{ccc}
                       1 & 1   & 0 \\
                       10 & -2 & -2 \\
                       -6 &0   & 0
                      \end{array}
                    \right),\quad\textrm{spec}(\mathcal{K}_{\,+1,\textsc{VAC}_3})=\{-1,-2\sqrt{3},2\sqrt{3}\},
\ee
and we expect  particular solutions in this case with the given leading order, however,  due to the irrational Kowalevskaya exponents the resulting series will contain logarithmic terms.


\section{Conclusion}
In this paper we have considered the possible singular behaviors and asymptotic limits of vacuum isotropic  universes in the fully quadratic gravity theory which apart from the Einstein term contains terms proportional to a linear combination of $R^2, \textrm{Ric}^2$ and $\textrm{Riem}^2$. Taking into account various asymptotic conditions that have to hold in order to have admissible solutions, we are left with only three possible asymptotic decompositions of the vacuum vector field near the singular state.

It turns out that a prominent role in the \emph{early} asymptotic evolution of both flat and curved vacua in this theory is played by a scaling form that behaves as $t^{1/2}$ near the initial singularity. Using various asymptotic and geometric arguments, we were able to built a solution of the field equations in the form of a Fuchsian formal series expansion compatible with all other constraints, dominated asymptotically to leading order by this solution and having the correct number of arbitrary constants that makes it a general solution of the field equations. In this way, we  conclude that this exact solution is an early time attractor of all homogeneous and isotropic vacua of the theory, thus proving stability against such `perturbations'.

For open vacua, there is a 2-parameter family of Fuchsian solutions that is dominated asymptotically by the Milne form both for past and future singularities. In the case of closed models, we have logarithmic solutions coming from a manifold of initial conditions with smaller dimension than the full phase space but dominated asymptotically by the same $a(t)\sim t$ form.

It is instructive to also comment on our present results in connection with results of \cite{kolionis1} on the stability of radiation, curved universes for the same class of theories. We have shown that at early times both radiation and vacuum, flat or curved universes are past-attracted by the `universal' $t^{1/2}$ asymptote and this is the most dominant feature in all these models. However, here we have shown that the behavior of open vacua in these theories is more complex.  For they allow novel types of asymptotic behaviour, universes that emerge from initial data sets of smaller dimension and valid for both early and late times. These universes asymptote to the Milne form during their early and late evolution toward singularities. Closed vacua, on the other hand, develop in time  as more complex solutions that are characterized by logarithmic formal series, but asymptotically their leading order is described  again by simple singularities similar to the open case treated here.

The existence of the Milne singularity and the attractor properties of our solutions bear a potential significance for the ekpyrotic scenario and its cyclic extension, wherein the passage through the singularity in these models, `the linchpin of the cyclic picture', depends on the stability of a Milne-type state under various kinds of perturbations \cite{khouri1, khouri2,st1,st2}. In particular, during the brane collision it is found that  spacetime asymptotes to Milne and so it is expected that higher-order derivative corrections will be small during such a phase, cf. \cite{toley,erickson,leh}. Our work indicates that such Milne states may indeed dynamically  emerge as stable asymptotes during the evolution in any theory with higher-order corrections in vacuum, or with a radiation content. What remains  is an interesting issue (that can be fully addressed with our asymptotic methods), that is  to find whether the `compactified Milne mod $\mathbb{Z}_2$'$\times\mathbb{R}_3$ space  monitoring the reversal phase in the ekpyrotic and cyclic scenarios also emerges asymptotically as a stable attractor in the dynamics of higher-order gravity, when the matter content is a fluid with a general equation of state.


\section*{Acknowledgements}
We are grateful to an anonymous referee for  useful suggestions.



\begin{thebibliography}{99}
\bibitem{ycb}
Y. Choquet-Bruhat, \emph{General Relativity and the Einstein Equations}, (OUP, 2009).
\bibitem{star}
A. A. Starobinski, \emph{Phys. Lett.} \textbf{B91} (1980) 99.
\bibitem{page}
D. N. Page, \emph{Phys. Rev.} \textbf{D36} (1987) 1607.
\bibitem{mijic-morris-suen}
M. B. Miji\'c, M. S. Morris and W. Suen, \emph{Phys. Rev.} \textbf{D39} (1989) 1511.
\bibitem{co-fl93b}
S. Cotsakis and G. P. Flessas, \emph{Phys. Lett.} \textbf{B319} (1993) 69.
\bibitem{ba-mid1}
J. D. Barrow and J. Middleton, \emph{Phys. Rev.} \textbf{D75} (2007) 123515.
\bibitem{ba-mid2}
J. Middleton and J. D. Barrow, \emph{Phys. Rev.} \textbf{D77} (2008) 103523.

\bibitem{mtw}
C.W. Misner, K.S. Thorne and J.A. Wheeler, \emph{Gravitation} (Freeman, San Francisco, 1973).
\bibitem{bar-ott-83}
J. D. Barrow, A. C. Ottewil,\emph{J. Phys. A: Math. Gen.} \textbf{16} 12  (1983), 2757-2776.
\bibitem{kolionis1}
S. Cotsakis, G. Kolionis and A. Tsokaros, 	\emph{Phys. Lett.} \textbf{B721} (2013) 1-6; arXiv:1211.5255.
\bibitem{CB}
S. Cotsakis and J. D. Barrow, \emph{J. Phys. Conf. Ser.} \textbf{68} (2007) 012004.
\bibitem{go}
A. Goriely, \emph{Integrability and Nonintegrability of Dynamical Systems}, (World Scientific, 2001).
\bibitem{me12}
S. Cotsakis,  \emph{Int. J. Mod. Phys}. \textbf{D23} (2013) 1330003;  arXiv:1212.6737.
\bibitem{bach+weyl}
H. Weyl, \emph{Ann. Phys. Leipzig} \textbf{59} (1919) 101; R. Bach, \emph{Math. Z.} \textbf{9} (1921) 110.
\bibitem{K1}
S. Cotsakis and A. Tsokaros, Phys. Lett. \textbf{B651} (2007) 341-344.
\bibitem{horizonless1}
A. Frenkel and K. Brecher, \emph{Phys. Rev.} \textbf{D26} (1982) 368; see also,
T. Rothman and P. Anninos, \emph{Phys. Rev.} \textbf{D44} (1991) 3087.
\bibitem{khouri1}
 J. Khoury, B. A. Ovrut, P. J. Steinhardt, N. Turok, \emph{Phys. Rev. } \textbf{D64} (2001) 123522.
\bibitem{khouri2}
J. Khoury, B. A. Ovrut, N. Seiberg, P. J. Steinhardt, N. Turok, \emph{Phys. Rev.} \textbf{D65} (2002) 086007.
\bibitem{st1}
P. J. Steinhardt, N. Turok, \emph{Phys. Rev.} \textbf{D65} (2002) 126003.
\bibitem{st2}
P. J. Steinhardt, N. Turok, \emph{Science} \textbf{296} (2002) 1436.
\bibitem{toley}
A. J. Tolley, N. Turok, P. J. Steinhardt, 	\emph{Phys. Rev.} \textbf{D69} (2004) 106005.
\bibitem{erickson}
J. K. Erickson, D. H. Wesley, P. J. Steinhardt, N. Turok, \emph{Phys. Rev.} \textbf{D69} (2004) 063514.
\bibitem{leh}
J. -L.  Lehners, 	\emph{Phys. Rept.} \textbf{ 465} (2008) 223.

\end{thebibliography}
\end{document}